# Chroma⁺Phy – A Living Wearable Connecting Humans and Their Environment

Theresa Schubert[1]

[1] Faculty of Media, Chair of Media Environments, Bauhaus-Universität Weimar, Germany

`theresa.schubert-minski@uni-weimar.de`

**Abstract.** This research presents an artistic project aiming to make cyberfiction become reality and exemplifying a current trend in art and science collaborations. Chroma⁺Phy is a speculative design for a living wearable that combines the protoplasmic structure of the amoeboid acellular organism *Physarum polycephalum* and the *chromatophores* of the reptile *Chameleon*. The underpinning idea is that in a future far away or close, on planet earth or in outer space, humans will need some tools to help them in their social life and day-to-day routine. Chroma⁺Phy enhances the body aiming at humans in extreme habitats for an aggression-free and healthy life. Our approach will address actual issues of scientific discovery for society and catalyse idea translation through art and design experiments at frontiers of science.

**Keywords:** body, design, wearable, hybrid, artificial vascularization, distributed sensing, decentralized intelligence, health, novel substrates.

## 1      Introduction

Let's imagine we live in outer space, somewhere far away from our orbit:

- There is no day or night-time anymore as the human body is used to. Chroma⁺Phy is showing night-time by lighter colours. Thus humans are able to follow their circadian rhythm and minimising risk of sleeping disorders and related health issues.
- UV-radiation from a sun will be very high. Human skin will need a measurement device to alert when radiation levels are getting too high by turning black. Physarum polycephalum is an excellent light sensor. It prefers darker environments thus it would try to move and change its physical pattern visibly.
- Chroma⁺Phy can measure temperature and air humidity and reacts with colours thus the wearer can take steps to counter dehydration.
- Chroma⁺Phy can sense the body temperature of the wearer and the heart beat through vibration. Accordingly it will make an interpretation of his current emotional state by changing colours, which will help deciding how or if at all to communicate with fellow humans.

## 2 Future Manual

Chroma+Phy comes in a box, ready to be applied directly onto the skin. It sits between a transparent layer of special medical adhesive silicone, that is permeable to air, humidity and temperature. After more than 48h of usage it has to be returned into the box that contains a nutritious, humid gel to 'recharge' for a few hours. Then it can be worn again, in total up to three or four weeks.

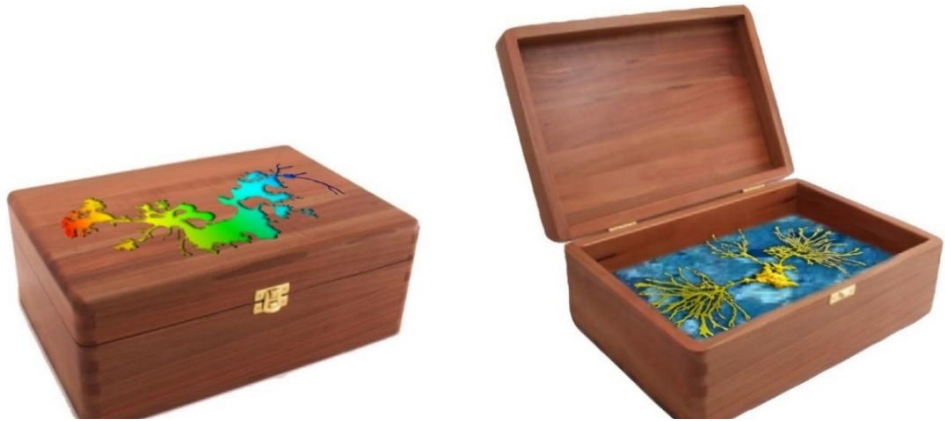

**Fig. 1.** Prototype of box containing the Chroma+Phy (left), box opened with living organism inside and recharge media (right)

## 3 Methods

The slime mould *Physarum polycephalum* has a complex life cycle. In its most active phase it looks like an amorphous yellowish mass with networks of protoplasmic tubes. The plasmodium behaves and moves as a giant amoeba. It is possible to divide the plasmodium and it will live on as separate entities or merge with another blob to one. It is a large single cell capable of distributed sensing and primitive memory [1]. The most prominent feature of the reptile *Chameleon* is its ability to change colours. Its function is in social signalling, in reactions to temperature and other environmental conditions. Colour change signals a chameleon's physiological condition and intentions to other chameleons. Chameleons have specialised cells, *chromatophores*, which contain pigments in their cytoplasm in three layers below their transparent outer skin. Dispersion of the pigment granules in the chromatophores sets the intensity of each colour, which can change due to rapid relocation of their particles of pigment [2].

We use this specific function of colour change combined with the decentralised logic of Physarum and its ability to attach and connect to any surface ignoring all gravitational laws. Inside the membrane Physarum is filled with cytoplasm. Through genetic manipulation Physarum's membrane and cytoplasm can be merged with Chameleon's chromatophores mechanism. For our research we designed a series of in vitro experiments that mark first steps towards the realisation at first without transgenic methods.

## 4    Discussion

We propose a hybrid organism that changes colour to indicate the intensity of the wearer's emotions, the percentage of outside humidity/radiation, the temperature as well as the circadian rhythm. The aim is to understand the human body and to improve communication when living in extreme environments. Currently still a speculative design, it exceeds pure fiction as a lot of experiments towards living wearable, control of Physarum polycephalum, chromatophores functionality, and novel silicone substrates can be made in real. Our aim is to make a transgenic Physarum in varying colours.

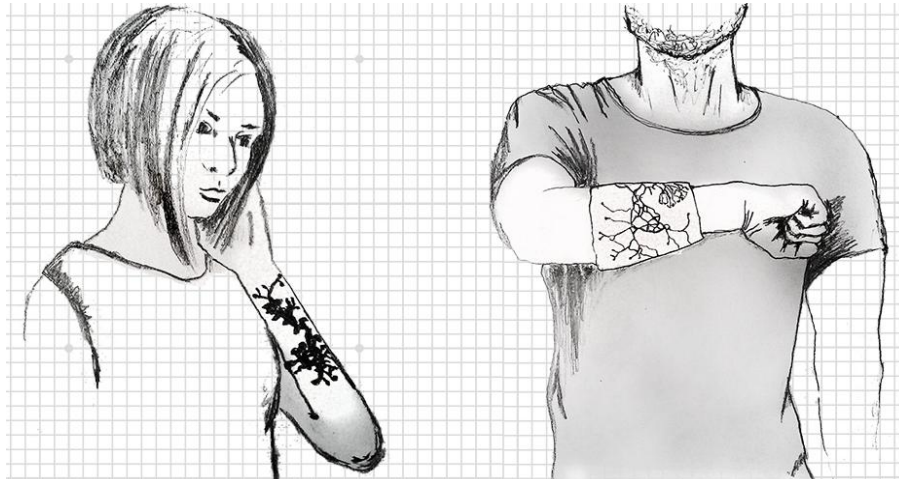

**Fig. 2.** Artist impression of Chroma$^+$Phy as it might be worn by users

In this context it is also interesting to analyse the implication of such a wearable in a future society. What does it mean if we live in a quasi-symbiosis with a hybrid organism? Would our behaviour change, will our human interactions improve or the opposite? Will we develop a new understanding and rapport to 'primitive' organisms?